\documentclass{ws-ijmpa}

\begin{document}

\markboth{A. V. Kuznetsov, N. V. Mikheev, and A. A. Okrugin}
{Reexamination of a Bound on the Dirac Neutrino Magnetic Moment} 

%%% Title: Reexamination of a Bound on the Dirac Neutrino Magnetic Moment from the 
%%% Supernova Neutrino Luminosity

%%%%%%%%%%%%%%%%%%%%% Publisher's Area please ignore %%%%%%%%%%%%%%
\catchline{}{}{}{}{}
%%%%%%%%%%%%%%%%%%%%%%%%%%%%%%%%%%%%%%%%%%%%%%%%%%%%%%%%%%%%%%%%%%%

\title{REEXAMINATION OF A BOUND \\ 
ON THE DIRAC NEUTRINO MAGNETIC MOMENT \\ 
FROM THE SUPERNOVA NEUTRINO LUMINOSITY} 

\author{\footnotesize A.V. KUZNETSOV, N.V. MIKHEEV and A.A. OKRUGIN }

\address
{Division of Theoretical Physics, Department of Physics,\\
Yaroslavl State University, Sovietskaya 14,\\
150000 Yaroslavl, Russian Federation\\
avkuzn@uniyar.ac.ru, mikheev@uniyar.ac.ru, okrugin@uniyar.ac.ru}

\maketitle

\pub{Received (Day Month Year)}{Revised (Day Month Year)}

\begin{abstract}
The neutrino helicity-flip process under the conditions of the supernova core 
is reinvestigated. Instead of the uniform ball model for the SN core used 
in previous analyses, realistic models for 
radial distributions and time evolution 
of physical parameters in the SN core are considered. A new upper bound on 
the Dirac neutrino magnetic moment is obtained from the limit on the 
supernova core luminosity for $\nu_R$ emission. 

\end{abstract}

\ccode{PACS numbers: 13.15.+g, 95.30.Cq, 97.60.Bw}

%%%%%%%%%%%%%%%%%%%%%%%%%%%%%%%%%%%%%%%%%%%%%%%%%%%%%%%%%%%%%%%%%%%%%%
\section{Introduction}
\label{sec:Introduction}
%%%%%%%%%%%%%%%%%%%%%%%%%%%%%%%%%%%%%%%%%%%%%%%%%%%%%%%%%%%%%%%%%%%%%%

Nonvanishing neutrino magnetic moment leads to the helicity-flip process 
where the left-handed neutrinos produced in the stellar interior
could convert into the right-handed neutrinos being sterile with respect to the weak 
interaction, and this can be important e.g. for the stellar 
energy-loss. 

A considerable interest to the neutrino magnetic moment arised after 
the great event of the $SN1987A$, 
in connection with the modelling of a supernova explosion, where 
a gigantic neutrino outflow defines in fact the process energetics. 
It means that such  a microscopic neutrino characteristic, as the neutrino 
magnetic moment, would have a crucial influence on macroscopic properties 
of these astrophysical events. 
Too large outflow of right-handed neutrinos, produced due to the magnetic 
moment interaction, from the core would leave no enough energy 
to explain the observed neutrino luminosity of the supernova. 
Thus, the upper bound on the neutrino magnetic moment 
can be established. 

The neutrino helicity flip $\nu_L \to \nu_R$ under physical conditions 
corresponding to the central region of a supernova 
has been studied in a number of works (see, e.g., Refs.~\refcite{Barbieri:1988}--\refcite{Ayala:2000}; 
a more extended reference list is given in Ref.~\refcite{Kuznetsov:2007}). 
The process is possible due to the interaction of 
the Dirac-neutrino magnetic moment with a virtual 
plasmon, which can be both generated and absorbed: 
\begin{eqnarray}
\nu_L \to \nu_R + \gamma^*, \quad \nu_L + \gamma^* \to \nu_R \, .
\label{eq:conversion}
\end{eqnarray}

In Ref.~\refcite{Barbieri:1988}, the neutrino helicity flip was described in terms of 
scattering by plasma electrons and protons ($\nu_L e^- \to \nu_R e^-$ 
and $\nu_L p \to \nu_R p$, respectively) in a supernova core 
immediately after the collapse. However, the important 
polarization effects of the plasma on the photon propagator 
were not considered in that work. Instead, the 
photon dispersion was taken into account phenomenologically 
by introducing the so-called thermal mass of 
a photon into the propagator. The above-mentioned 
effects were considered more consistently in Refs.~\refcite{Ayala:1999,Ayala:2000},  
where the effect of a high-density astrophysical plasma 
on the photon propagator was taken into account using 
the thermal field-theory formalism. However, an analysis 
of works~\cite{Ayala:1999,Ayala:2000} showed that they concerned only 
the electron component of the plasma, namely, only the 
channel $\nu_L e^- \to \nu_R e^-$, and only the electron contribution 
to the photon propagator, whereas the proton component 
of the plasma was not considered at all. This 
seemed to be even stranger because the plasma-electron 
and proton contributions to the neutrino spin flip were estimated 
earlier~\cite{Barbieri:1988} to be of the same order.

A detailed analysis of the processes~(\ref{eq:conversion}), 
with neutrino-helicity conversion due to the interaction with both 
plasma electrons and protons via a virtual plasmon and 
with taking account of polarization effects of the plasma 
on the photon propagator was performed in Ref.~\refcite{Kuznetsov:2007}. 
In particular, 
according to the numerical analysis, the contribution 
of the proton component of the plasma was not 
only significant, but even dominant. 

However, all those analyses~\cite{Barbieri:1988}\cdash\cite{Kuznetsov:2007} 
were based on a very simplified model of the supernova core as the uniform ball with some
averaged values of physical parameters. 
Moreover, the parameter values look, in modern views, rather too high than typical. 
It should be mentioned also that the improvement of the 
bound of Refs.~\refcite{Ayala:1999,Ayala:2000} with respect to the bound 
of Ref.~\refcite{Barbieri:1988} 
was based in part on the enlargement by the factor of 2 of the supernova 
core volume, 
% made in Refs.~\refcite{Ayala:1999,Ayala:2000} if compared with 
% Ref.~\refcite{Barbieri:1988}, 
while the core density was taken to be the same, 
$\rho_c \simeq 8 \times 10^{14} \ {\rm g \ cm}^{-3}$. This means that the 
core mass appeared to be in Refs.~\refcite{Ayala:1999,Ayala:2000} of the order 
of $3 \, M_{\odot}$, which is nearly twice as large as the mass of the 
supernova remnant believed to be typical.

The aim of this paper is to 
make the estimation of the Dirac neutrino magnetic moment from the limit on the supernova 
core luminosity for $\nu_R$ emission by a more consistent way, taking some
radial distributions and time evolution of physical parameters from some realistic models of
the supernova core.

For completeness, we consider here a general case of the magnetic moment matrix 
$\mu_{\nu_i \nu_j} \equiv \mu_{i j}$ 
(i.e. both diagonal and transition magnetic moments), where $\nu_i, \, \nu_j$ are the neutrino 
mass eigenstates. The neutrino states $\nu_\ell$ with definite flavors $\ell$ created in weak processes
are the superpositions of the neutrino mass eigenstates:
\begin{eqnarray}
\nu_\ell = \sum\limits_i U_{\ell i}^* \nu_i \, ,
\label{eq:superposition}
\end{eqnarray}
where $U_{\ell i}$ is the unitary leptonic mixing matrix by Pontecorvo--Maki--Nakagawa--Sakata. 
It means that the value of the magnetic moment squared in all equations of 
Ref.~\refcite{Kuznetsov:2007} should be considered as an effective value. 
For the processes with the initial electron neutrino one should replace
\begin{eqnarray}
\mu_\nu^2 \to \mu_{\nu_e}^2 \equiv \sum\limits_i 
\left | \sum\limits_j \mu_{i j} U_{e j} \right |^2 \, ,
\label{eq:munu_e_eff}
\end{eqnarray}
and similarly for the muon and tau initial neutrinos. 

The paper is organized as follows. In Sec.~\ref{sec:Illustration} we give 
a clear illustration of the fact that neutrino scattering 
by protons dominates over their scattering by 
plasma electrons, basing on an analysis of a simplified case  
of the completely degenerate plasma, $T = 0$. 
Sec.~\ref{sec:Uniform} contains a summary of the procedure~\cite{Kuznetsov:2007} 
of obtaining the upper bound on the electron-neutrino 
magnetic moment from the $SN1987A$ data, in the frame of the uniform ball model  
for the supernova core. 
In Sec.~\ref{sec:Models} we make the estimation by a more reliable way, with taking 
account of radial distributions and time evolution of physical parameters, from realistic 
models of the SN core. The upper bounds are obtained on the combination of 
the effective magnetic moments 
of the electron, muon and tau neutrinos from the condition of not-spoiling the subsequent 
Kelvin---Helmholtz stage of the supernova explosion by emission of right-handed neutrinos 
during a few seconds after the collapse. 
% 

%%%%%%%%%%%%%%%%%%%%%%%%%%%%%%%%%%%%%%%%%%%%%%%%%%%%%%%%%%%%%%%%%%%%%%
\section{Illustration: completely degenerate plasma at $T = 0$}
\label{sec:Illustration}
%%%%%%%%%%%%%%%%%%%%%%%%%%%%%%%%%%%%%%%%%%%%%%%%%%%%%%%%%%%%%%%%%%%%%%

The comparison of the typical parameters of the supernova core, where the 
temperature is believed to be of order 
$T \simeq \ $ 15--30 MeV, while the electron and neutrino 
chemical potentials are $\eta_e \simeq \ $ 200--250 MeV 
and $\eta_{\nu_e} \simeq$ 100 MeV, respectively, shows that the temperature is 
the smallest physical parameter.
\footnote{Hereafter we consider neutrinos as a quasiequilibrium gas described by 
the distribution functions $f_\nu (T, \eta_{\nu_e})$ for the 
electron neutrinos and $f_\nu (T, 0)$ for the muon and tau neutrinos. 
This is believed to be a rather good approximation inside the SN core 
during a few seconds after the collapse.} 
Thus, the limiting 
case of the completely degenerate plasma, $T = 0$, seems 
to give a reasonable estimate. It is remarkable that for the 
zero temperature limit the contributions from neutrino 
scattering by protons and electrons to the neutrino creation 
probability can be evaluated analytically using 
Eqs. (20) and (21) and the corresponding formulas 
from Appendix A of Ref.~\refcite{Kuznetsov:2007}. 

It is appropriate to analyse the function $\Gamma_{\nu_R} (E)$ defining the
energy spectrum of right-handed neutrinos. 
In other words, this function specifies the 
number of right-handed neutrinos emitted per 1 MeV of 
the neutrino energy spectrum per unit time from unit 
volume of the central region of a supernova: 

\begin{eqnarray}
\frac{\mathrm{d} n_{\nu_R}}{\mathrm{d} E} = 
\frac{E^2}{2 \, \pi^2} \, \Gamma_{\nu_R} (E) \,.
\label{eq:dn/dE}
\end{eqnarray}

The contribution of ultrarelativistic electrons to the function 
$\Gamma_{\nu_R} (E)$ in the case $T = 0$ can be obtained from the 
above-mentioned formulas of Ref.~\refcite{Kuznetsov:2007} in the simple form:

\begin{equation}
\Gamma_{\nu_R}^{(e)}(E) = \frac{\mu_{\nu_e}^2\,m_\gamma^2}{2 \, \pi}\,
(\eta_{\nu_e} - E) \, \theta (\eta_{\nu_e} - E)\,,
\label{eq:zerotlimit_Le}
\end{equation}
where $E$ is the right-handed 
neutrino energy, $\mu_{\nu_e}$ is the effective electron neutrino 
magnetic moment~(\ref{eq:munu_e_eff}), 
$m_\gamma^2 = 2 \, \alpha \, \eta_e^2/\pi$ is the squared mass 
of a transverse plasmon at $T = 0$, and $\theta (x)$ is the step function. 
% and $\eta_{\nu_e}$ is the electron neutrino chemical potential.  

The analytical expression describing the proton contribution 
turns out to be more complicated since it 
depends additionally on the proton mass. The plasma 
charge neutrality condition for $T = 0$ takes the form $n_p = n_{e^-}$ 
and ensures that the electron and proton Fermi
momenta are the same: $k^{(e)}_{\rm{F}} = k^{(p)}_{\rm{F}}$. Then, the proton 
chemical potential coinciding with the Fermi energy is
$\eta_p = E^{(p)}_{\rm{F}} =
\sqrt{m_p^2 + \eta_e^2}$ and the proton contribution is 
expressed in terms of the proton Fermi velocity 
$v_{\rm{F}} 
= {k^{(p)}_{\rm{F}}}/{E^{(p)}_{\rm{F}}} 
= {\eta_e}/{\eta_p} 
= {\eta_e}/{\sqrt{m_p^2 + \eta_e^2}}$.
As a result, the proton contribution can be expressed in the form: 
\begin{equation}
\Gamma_{\nu_R}^{(p)}(E) = 
\frac{\mu_{\nu_e}^2\,m_\gamma^2\,\eta_{\nu_e}}{2\,\pi} 
\, \varphi_p (y) \,, \quad
y = \frac{E}{\eta_{\nu_e}} \,, \quad 0 \leqslant y \leqslant 1 \,.
\label{eq:zl_defns} 
\end{equation}
Here, the function $\varphi_p (y)$ has the different forms in two 
intervals: it is  
\begin{equation}
\varphi_p (y) = \frac{1+v_{\rm{F}}/3}{1-v_{\rm{F}}} \; y \,,  
\label{eq:f_y_protons1}
\end{equation}
for $0 \leqslant y \leqslant (1-v_{\rm{F}})/(1+v_{\rm{F}})$, and 
\begin{equation}
\varphi_p (y) = \frac{1-y}{v_{\rm{F}}} 
\left[1 - \frac{(1-v_{\rm{F}})^2}{12 \, y^2\, v_{\rm{F}}}\,(1-y)\,(1+2\,y) \right]\,,
\label{eq:f_y_protons2}
\end{equation}
for 
$(1-v_{\rm{F}})/(1+v_{\rm{F}}) \leqslant y \leqslant 1$.

Note that the formal turn to the limit $m_p \to 0$,
i.e. $v_{\rm{F}} \to 1$, in Eqs.~(\ref{eq:zl_defns})--(\ref{eq:f_y_protons2}) 
yields $\varphi_p (y) \to \varphi_e (y) = (1-y)\; \theta(1-y)$, where the function 
$\varphi_e (y)$ can be introduced in 
Eq.~(\ref{eq:zerotlimit_Le}) in complete analogy with Eq.~(\ref{eq:zl_defns}). Thus, as 
expected, Eq.~(\ref{eq:zerotlimit_Le}) for the electron contribution is reproduced. 

Figure 1 
shows the plots of the function $\varphi_p (y)$ for 
$v_{\rm{F}}$ = 1, $v_{\rm{F}}$ = 0.394, and $v_{\rm{F}}$ = 0. 
The value $v_{\rm{F}}$ = 0.394 corresponds 
to the effective proton mass $m_p \simeq 700$ MeV in a plasma 
with a nuclear density $3 \times 10^{14} \ {\rm g \ cm}^{-3}$ 
(see Ref.~\refcite{Raffelt:1996}, p. 152). The value $v_{\rm{F}} = 0$ corresponds 
to the formal limit $m_p \to \infty$ 
for which this function is also significantly 
simplified: $\varphi_p (y) \to \varphi_{\infty} (y) = y\; \theta(1-y)$. 

%%%%%%%%%%%%%%%%%%%%%%%%%%%%%%%%%%%%%%%%%%%%%%%%%%%%%%%%%%%%%%%%%%%%%%
\begin{figure}
\begin{center}
\includegraphics*[width=0.75\textwidth]{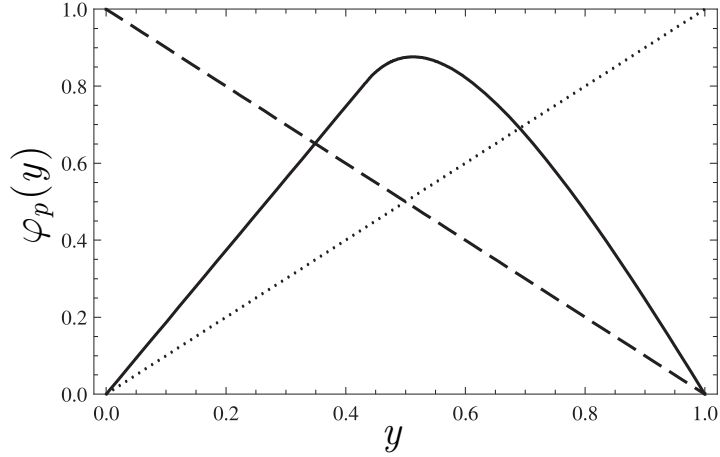}
\caption{Plots of the function $\varphi_p (y)$ for various $v_{\rm{F}}$ values. The 
dependence $\varphi_e (y) = (1-y)$ for the electron contribution is 
reproduced for $v_{\rm{F}} = 1$ (dashed line). The value $v_{\rm{F}} = 0.394$ 
(solid curve) corresponds to the effective proton mass $m_p \simeq 700$
MeV. The case $v_{\rm{F}} = 0$ (dotted line) corresponds to the 
limit of infinitely large proton mass.} 
\end{center}
\label{fig:phi_y}
\end{figure}
%%%%%%%%%%%%%%%%%%%%%%%%%%%%%%%%%%%%%%%%%%%%%%%%%%%%%%%%%%%%%%%%%%%%%%

The function $\Gamma_{\nu_R} (E)$ defined in Eq.~(\ref{eq:dn/dE}) 
determines as well the
right-handed neutrino emissivity of a supernova core, i.e. the energy 
passed away by right-handed neutrinos per 1 MeV of the neutrino energy spectrum 
per unit time from unit volume: 

\begin{eqnarray}
Q_{\nu_R}
 = E \, \frac{\mathrm{d} n_{\nu_R}}{\mathrm{d} E} = 
\frac{E^3}{2 \, \pi^2} \, \Gamma_{\nu_R} (E) \,.
\label{eq:Q}
\end{eqnarray}

According to Eqs.~(\ref{eq:dn/dE}) and~(\ref{eq:Q}), 
the right-handed neutrino emissivity is given by the formula 
\begin{eqnarray}
Q_{\nu_R}
 = \frac{\mu_{\nu_e}^2\,m_\gamma^2\,\eta_{\nu_e}^4}{4\,\pi^3} 
 \, y^3  
 \left[ \varphi_e (y) + \varphi_p (y) \right] \,.
\label{eq:Q2}
\end{eqnarray}
%

%%%%%%%%%%%%%%%%%%%%%%%%%%%%%%%%%%%%%%%%%%%%%%%%%%%%%%%%%%%%%%%%%%%%%%
\begin{figure}
\begin{center}
\includegraphics*[width=0.75\textwidth]{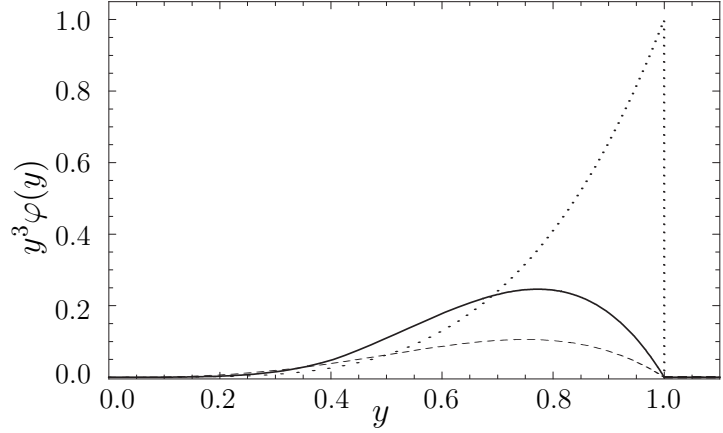}
\caption{The function $y^3 \, \varphi (y)$ defining the contributions from 
electrons (dashed line) and protons with $m_p \simeq 700$ MeV (solid line) and 
$m_p \to \infty$ (dotted line) to the right-handed neutrino emissivity 
at $T = 0$.} 
\end{center}
\label{fig:phi_y2}
\end{figure}
%%%%%%%%%%%%%%%%%%%%%%%%%%%%%%%%%%%%%%%%%%%%%%%%%%%%%%%%%%%%%%%%%%%%%%

The difference between the electron and proton contributions 
to the quantity given by Eq.~(\ref{eq:Q2}) is illustrated in 
Fig.~\ref{fig:phi_y2}. It is clearly seen that the factor $y^3$
causes the increasing of the proton contribution to the emissivity. 

%%%%%%%%%%%%%%%%%%%%%%%%%%%%%%%%%%%%%%%%%%%%%%%%%%%%%%%%%%%%%%%%%%%%%%
\section{Uniform ball model for the SN core}
\label{sec:Uniform}
%%%%%%%%%%%%%%%%%%%%%%%%%%%%%%%%%%%%%%%%%%%%%%%%%%%%%%%%%%%%%%%%%%%%%%

The spectral density of the supernova core luminosity via 
right-handed neutrinos is defined by the function 
$\Gamma_{\nu_R} (E)$ as follows: 

\begin{eqnarray}
\frac{\mathrm{d} L_{\nu_R}}{\mathrm{d} E}
 = V\, \frac{\mathrm{d} n_{\nu_R}}{\mathrm{d} E} \, E = 
V \, \frac{E^3}{2 \, \pi^2} \, \Gamma_{\nu_R} (E) = 
V \, \frac{\mu_{\nu_e}^2\,m_\gamma^2\,\eta_{\nu_e}^4}{4\,\pi^3} 
 \, y^3 \, \varphi^{\mbox{\scriptsize (num)}} (y, T) \,.
\label{eq:dL/dE}
\end{eqnarray}

%%%%%%%%%%%%%%%%%%%%%%%%%%%%%%%%%%%%%%%%%%%%%%%%%%%%%%%%%%%%%%%%%%%%%%
\begin{figure}
\begin{center}
\includegraphics*[width=0.75\textwidth]{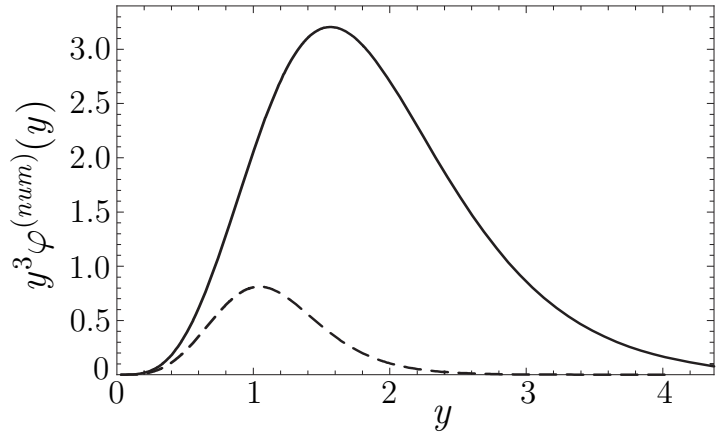}
\caption{The function $y^3 \, \varphi^{\mbox{\scriptsize (num)}} (y, T)$ representing 
the result of the numerical calculation of 
the right-handed neutrino emissivity at $T = 30$ MeV (dashed line) and 
$T = 60$ MeV (solid line).}
\end{center}
\label{fig:emissivity}
\end{figure}
%%%%%%%%%%%%%%%%%%%%%%%%%%%%%%%%%%%%%%%%%%%%%%%%%%%%%%%%%%%%%%%%%%%%%%

Here, $V$ is the volume of the neutrino-emitting region,
$m_\gamma$ is the mass of a transverse plasmon, 
\begin{equation}
m_\gamma^2 = \frac{2 \, \alpha}{\pi} \left({\eta_e}^2 + 
\frac{\pi^2 T^2}{3} \right).
\label{eq:m_gamma}
\end{equation}
The function $\varphi^{\mbox{\scriptsize (num)}} (y, T)$ 
introduced in Eq.~(\ref{eq:dL/dE})
similarly to Eqs.~(\ref{eq:zl_defns}) and~(\ref{eq:Q2}) 
can be extracted from Ref.~\refcite{Kuznetsov:2007}. 
The function $y^3 \, \varphi^{\mbox{\scriptsize (num)}} (y, T)$ 
is plotted in Fig.~\ref{fig:emissivity} for 
two values of the averaged temperature and for the electron and 
electron-neutrino chemical 
potentials $\eta_e \simeq$ 300 MeV and $\eta_{\nu_e} \simeq$ 160 MeV. 
We neglected in our analysis~\cite{Kuznetsov:2007} the contributions of the 
processes with the initial muon and tau neutrinos. However, as will be shown 
below, these contributions appear to be essential. 

A comparison of Figs.~\ref{fig:phi_y2} and~\ref{fig:emissivity} shows that taking  
of a nonzero temperature leads to a shift of the maximum 
of the energy distribution of the luminosity 
towards higher energies of right-handed neutrinos. This 
additionally enhances the proton contribution. 

As a result, 
using the data on supernova $SN1987A$, a new astrophysical 
limit was imposed~\cite{Kuznetsov:2007} on the electron-neutrino 
magnetic moment: 
\begin{eqnarray}
\mu_\nu < (0.7-1.5) \, \times 10^{-12} \, \mu_{\rm B}\,.
\label{eq:mu_fr_Q}
\end{eqnarray}
This is a factor of two better than the previous 
constraint.\cite{Ayala:1999,Ayala:2000} 
We have to remind, however, that both the previous and this improved bound 
on the electron-neutrino magnetic moment were based 
on a very simplified model of the supernova core as the uniform ball with some
averaged values of physical parameters. In addition, the parameter values were 
set too high. 
For example, the upper limit $1.5 \, \times 10^{-12} \, \mu_{\rm B}$ 
in Eq.~(\ref{eq:mu_fr_Q}) corresponds to the SN core temperature 30 MeV, 
while the limit $0.7 \, \times 10^{-12} \, \mu_{\rm B}$ 
corresponds to the temperature 60 MeV. As is seen from Fig.~\ref{fig:emissivity}, 
the right-handed neutrino emissivity grows with temperature very rapidly. 
However, according to recent simulations of the SN explosion, the temperature 
values inside the SN core are believed not to exceed 40 MeV, see e.g. Fig. 4. 
Anyway, taking account of the radial distribution of physical parameters 
inside the SN core would give more solid results.

%%%%%%%%%%%%%%%%%%%%%%%%%%%%%%%%%%%%%%%%%%%%%%%%%%%%%%%%%%%%%%%%%%%%%%
\section{Models of the supernova core with radial 
distributions of physical parameters}
\label{sec:Models}
%%%%%%%%%%%%%%%%%%%%%%%%%%%%%%%%%%%%%%%%%%%%%%%%%%%%%%%%%%%%%%%%%%%%%%

In this section we make the estimation of the upper bound on the Dirac neutrino 
magnetic moment by a more reliable way, with taking account of radial 
distributions and time dependences of physical parameters from realistic models 
of the SN core. Here we consider the models in the inverse chronology.

%%%%%%%%%%%%%%%%%%%%%%%%%%%%%%%%%%%%%%%%%%%%%%%%%%%%%%%%%%%%%%%%%%%%%%
\subsection{The recent model of the O-Ne-Mg core collapse SN} 
\label{sec:Janka:2009}
%%%%%%%%%%%%%%%%%%%%%%%%%%%%%%%%%%%%%%%%%%%%%%%%%%%%%%%%%%%%%%%%%%%%%%

The very recent model was developed by H.-Th. Janka with collaborators who presented us 
the results of their simulations~\cite{Janka:2009} of the O-Ne-Mg core collapse
supernovae which were a 
continuation of model simulations of Refs.~\refcite{Kitaura:2006,Janka:2008}. 
The successful explosion results for this case have recently been
independently confirmed by the Arizona/Princeton SN modelling 
group,\cite{Dessart:2006,Burrows:2007} which found very similar results. 
So we were provided with a model whose explosion behavior was 
comparatively well understood and generally accepted.

We redefine Eq.~(\ref{eq:dL/dE}), where, instead of multiplying by the volume of 
the neutrino-emitting region $V$, we integrate over this volume to obtain 
the spectral density of the energy luminosity of a supernova 
core via right-handed neutrinos: 

\begin{eqnarray}
\frac{\mathrm{d} L_{\nu_R}}{\mathrm{d} E} = 
\int \, \mathrm{d} V  \, \frac{E^3}{2 \, \pi^2} \, \Gamma_{\nu_R} (E) \, .
\label{eq:dL/dE_int}
\end{eqnarray}

Here, taking the values defined in Eqs. (20) and (21) and the corresponding formulas 
from Appendix A of Ref.~\refcite{Kuznetsov:2007}, we take account of their 
dependence on the radius $R$ and time $t$. A comprehensive set of parameter  
distributions used in our estimation includes the profiles~\cite{Janka:2009} 
of the density $\rho$, the temperature $T$, the electron fraction $Y_e$, 
the fractions of electron neutrinos $Y_{\nu_e}$, electron anti-neutrinos 
$Y_{\bar\nu_e}$, and the fractions $Y_{\nu_x}$ for one kind of heavy-lepton neutrino 
or antineutrino ($\nu_x = \nu_{\mu, \tau}, \bar\nu_{\mu, \tau}$), which are treated 
identically. The time evolution of the parameter distributions is 
calculated~\cite{Janka:2009} within the interval until $\sim$ 2 sec after the bounce. 
For the sake of illustration, we present in 
Figs.~4--6
the radial distributions within the SN core, from 0 to 20 km, at the moment 
$t = 1.0$ sec after the bounce for the temperature,\cite{Janka:2009} for the chemical potentials 
of electrons $\eta_e$ and electron neutrinos $\eta_{\nu_e}$ 
(calculated on the base of the data of Ref.~\refcite{Janka:2009}), 
and for the proton nonrelativistic chemical potential 
$\eta_p^* = \eta_p - m_N^*$ defining the degeneracy of protons 
(calculated on the base of the data of Ref.~\refcite{Janka:2009} and 
of the effective nucleon mass $m_N^*$ in plasma, 
see Ref.~\refcite{Raffelt:1996}, p. 152).  

%%%%%%%%%%%%%%%%%%%%%%%%%%%%%%%%%%%%%%%%%%%%%%%%%%%%%%%%%%%%%%%%%%%%%%
\begin{figure}
\begin{center}
\includegraphics*[width=0.75\textwidth]{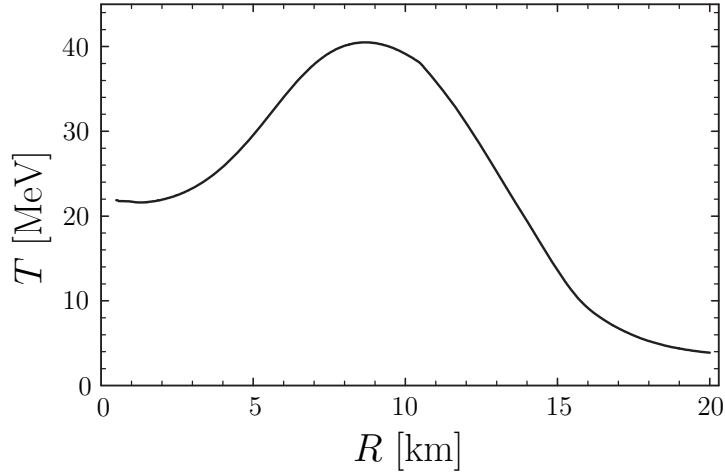}
\caption{The radial distribution for the temperature within the SN core at the moment 
$t = 1.0$ sec after the bounce, Ref. 6.}
\end{center}
\label{fig:T_R}
\end{figure}
%%%%%%%%%%%%%%%%%%%%%%%%%%%%%%%%%%%%%%%%%%%%%%%%%%%%%%%%%%%%%%%%%%%%%%

%%%%%%%%%%%%%%%%%%%%%%%%%%%%%%%%%%%%%%%%%%%%%%%%%%%%%%%%%%%%%%%%%%%%%%
\begin{figure}
\begin{center}
\includegraphics*[width=0.75\textwidth]{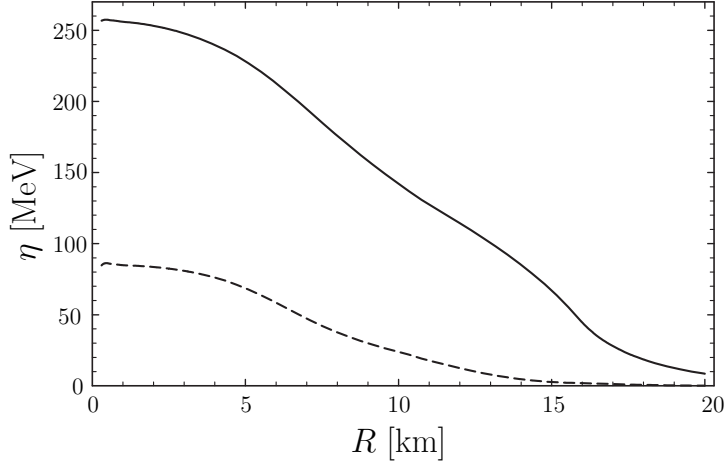}
\caption{The radial distributions for the chemical potentials 
of electrons (solid line) and electron neutrinos (dashed line) 
within the SN core at the moment $t = 1.0$ sec after the bounce.}
\end{center}
\label{fig:mue_R}
\end{figure}
%%%%%%%%%%%%%%%%%%%%%%%%%%%%%%%%%%%%%%%%%%%%%%%%%%%%%%%%%%%%%%%%%%%%%%

%%%%%%%%%%%%%%%%%%%%%%%%%%%%%%%%%%%%%%%%%%%%%%%%%%%%%%%%%%%%%%%%%%%%%%
\begin{figure}
\begin{center}
\includegraphics*[width=0.75\textwidth]{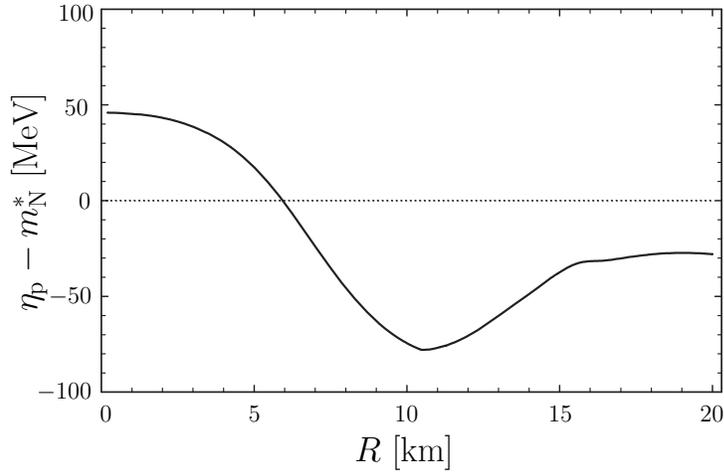}
\caption{The radial distribution for the proton nonrelativistic chemical potential 
$\eta_p^* = \eta_p - m_N^*$ within the SN core 
at the moment $t = 1.0$ sec after the bounce.}
\end{center}
\label{fig:mup_R}
\end{figure}
%%%%%%%%%%%%%%%%%%%%%%%%%%%%%%%%%%%%%%%%%%%%%%%%%%%%%%%%%%%%%%%%%%%%%%

To analyse the 
influence of the right-handed neutrino emission on the SN energy loss, we also used 
the time evolution of the total luminosity of all species of left-handed 
neutrinos,\cite{Janka:2009} presented in 
Fig.~7.

%%%%%%%%%%%%%%%%%%%%%%%%%%%%%%%%%%%%%%%%%%%%%%%%%%%%%%%%%%%%%%%%%%%%%%
\begin{figure}
\begin{center}
\includegraphics*[width=0.75\textwidth]{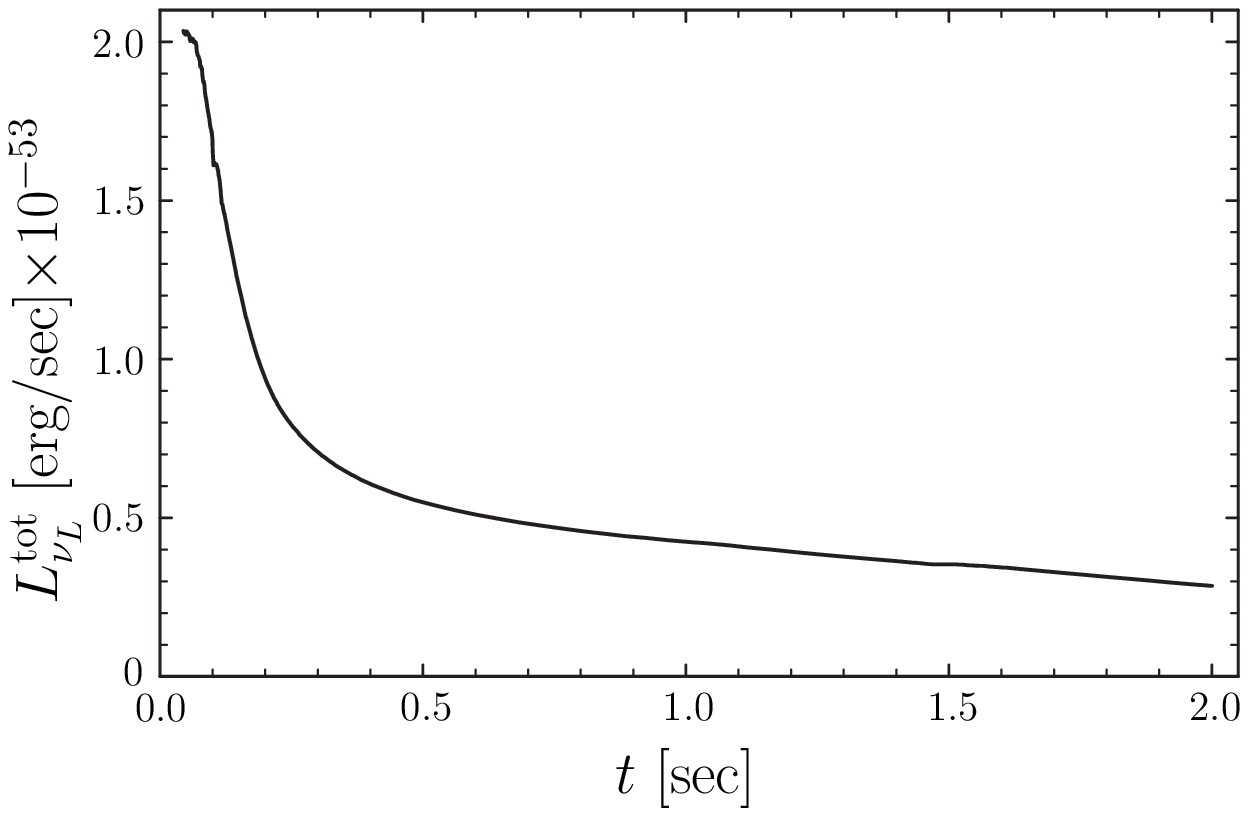}
\caption{The time evolution of the total luminosity of all active 
neutrino species, Ref. 6.} 
\end{center}
\label{fig:L_nuL_t}
\end{figure}
%%%%%%%%%%%%%%%%%%%%%%%%%%%%%%%%%%%%%%%%%%%%%%%%%%%%%%%%%%%%%%%%%%%%%%

Integrating Eq.~(\ref{eq:dL/dE_int}) over the neutrino energy, one obtains 
the time evolution of the right-handed neutrino luminosity:
\begin{eqnarray}
L_{\nu_R} (t) = \frac{1}{2 \, \pi^2} \, \int \, \mathrm{d} V  
\, \int\limits_0^\infty \, \mathrm{d} E \, E^3 \, \Gamma_{\nu_R} (E) \,.
\label{eq:L_def}
\end{eqnarray}
This is a novel cooling agent which would have to compete with the energy-loss 
via active neutrino species in order to affect the total cooling time scale 
significantly. Therefore, the observed $SN1987A$ signal duration indicates that 
a novel energy-loss via right-handed neutrinos is bounded by
\begin{eqnarray}
L_{\nu_R} < L_{\nu_L} \,.
\label{eq:E_lim}
\end{eqnarray}
Within the considered time interval until 2 sec after the bounce, 
one obtains from Eqs.~(\ref{eq:L_def}),~(\ref{eq:E_lim}) 
the time-dependent upper bound on the combination of the effective magnetic moments 
of the electron, muon and tau neutrinos. 
Assuming for simplicity that these effective magnetic moments 
are equal, one obtains the time evolution of 
the upper bound on some flavor-averaged neutrino magnetic moment $\bar \mu_\nu$
shown in Fig.~8, 
where $\bar \mu_{12} = \bar \mu_\nu/(10^{-12} \, \mu_{\rm B})$. 
 
%%%%%%%%%%%%%%%%%%%%%%%%%%%%%%%%%%%%%%%%%%%%%%%%%%%%%%%%%%%%%%%%%%%%%%
\begin{figure}
\begin{center}
\includegraphics*[width=0.75\textwidth]{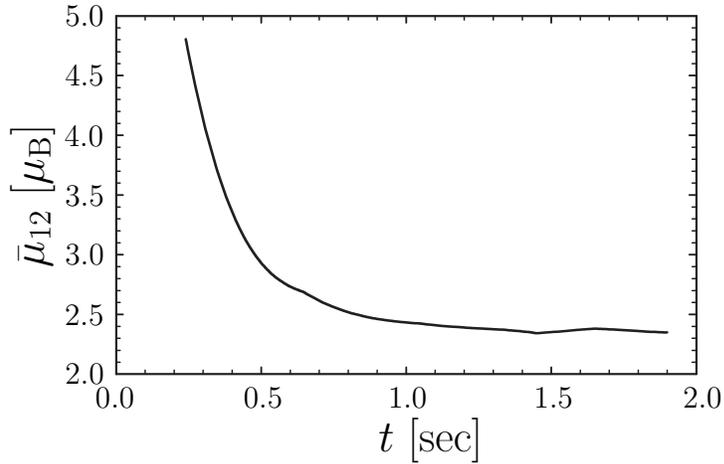}
\caption{The time evolution of the upper bound on the neutrino magnetic moment 
within the time interval until 2 sec after the bounce 
(in assumption that the effective magnetic moments of electron, muon and tau 
neutrinos are equal).} 
\end{center}
\label{fig:mu_nu_t}
\end{figure}
%%%%%%%%%%%%%%%%%%%%%%%%%%%%%%%%%%%%%%%%%%%%%%%%%%%%%%%%%%%%%%%%%%%%%%

As is seen from 
Fig.~8, 
the averaged upper bound tends to some value, providing the limit 
\begin{eqnarray}
\bar \mu_\nu < 2.4 \, \times 10^{-12} \, \mu_{\rm B}\,.
\label{eq:mu_lim_Janka09}
\end{eqnarray}
In a general case the combined limit on the effective magnetic moments 
of the electron, muon and tau neutrinos is
\begin{eqnarray}
\left [ \mu_{\nu_e}^2 + 0.71 \left(\mu_{\nu_\mu}^2 + \mu_{\nu_\tau}^2 \right) \right]^{1/2} 
< 3.7 \, \times 10^{-12} \, \mu_{\rm B}\,,
\label{eq:mu_lim_Janka09_comb}
\end{eqnarray}
where the effective magnetic moments are defined according to Eq.~(\ref{eq:munu_e_eff}). 
This limit is less stringent than the bound~(\ref{eq:mu_fr_Q}) obtained in the frame 
of the uniform ball model for the SN core, but it is surely more reliable. 
Additionally, the upper bound on the effective magnetic moments 
of muon and tau neutrinos is established. 

%%%%%%%%%%%%%%%%%%%%%%%%%%%%%%%%%%%%%%%%%%%%%%%%%%%%%%%%%%%%%%%%%%%%%%
\subsection{Earlier models of the SN explosion} 
\label{sec:other}
%%%%%%%%%%%%%%%%%%%%%%%%%%%%%%%%%%%%%%%%%%%%%%%%%%%%%%%%%%%%%%%%%%%%%%

The similar procedure of evaluation was performed with using of the data of 
the model\cite{Buras:2006} by R. Buras et al. (2006) 
of the two-dimensional hydrodynamic core-collapse 
supernova simulation for a 15 $M_\odot$ star. Namely, the radial distributions 
of parameters at the moments $t = 0.2, 0.4, 0.6, 0.8$ sec after the bounce 
in the model \textit{s15Gio\_32.a} were taken from Fig. 40 of Ref.~\refcite{Buras:2006}. 
Additionally, the fraction of electron neutrinos was evaluated as 
$Y_{\nu_e} \simeq (1/5) \, Y_e$. 
Calculating the right-handed neutrino luminosity with those parameters and 
putting the limit~(\ref{eq:E_lim}), 
where the total luminosity via active neutrino species $L_{\nu_L}$ in that model 
can be taken from Fig. 42 of Ref.~\refcite{Buras:2006}, 
one obtains that the upper bound on the flavor-averaged neutrino magnetic moment 
$\bar \mu_\nu$ also varies in time as in the previous case. 
The time-averaged upper bound on $\bar \mu_\nu$ corresponding 
to the interval 0.4--0.8 sec, is:
\begin{eqnarray}
\bar \mu_\nu < 2.7 \, \times 10^{-12} \, \mu_{\rm B}\,,
\label{eq:mu_lim_Buras06}
\end{eqnarray}
to be compared with the limit~(\ref{eq:mu_lim_Janka09}). 

Using the results of Ref.~\refcite{Pons:1999} by J.A. Pons et al. (1999) where 
the thermal and chemical evolution during the Kelvin-Helmholtz phase of the birth of a
neutron star was studied, taking the data from Figs. 9 and 14, we have obtained 
the time-averaged upper bound on $\bar \mu_\nu$ for the time interval 1--10 sec of 
the post-bounce evolution in the form:
\begin{eqnarray}
\bar \mu_\nu < 1.2 \, \times 10^{-12} \, \mu_{\rm B}\,.
\label{eq:mu_lim_Pons99}
\end{eqnarray}

We also used the results of Ref.~\refcite{Keil:1995} by W. Keil and H.-Th. Janka (1995) where 
the numerical simulations were performed of the neutrino-driven deleptonization and 
cooling of newly formed, hot, lepton-rich neutron star. Using the data presented 
in Figs. 3-9 on the SBH model (of the hot star with a ``small'' barionic mass), 
we have evaluated 
the time-averaged upper bound on $\bar \mu_\nu$ for the time interval 0.5--5 sec 
after the bounce in the form:
\begin{eqnarray}
\bar \mu_\nu < 1.1 \, \times 10^{-12} \, \mu_{\rm B}\,.
\label{eq:mu_lim_Keil95}
\end{eqnarray}

One can summarize that the upper bound on the flavor- and time-averaged 
neutrino magnetic moment at the Kelvin-Helmholtz phase of the supernova 
explosion occurs to be
\begin{eqnarray}
\bar \mu_\nu < (1.1 - 2.7) \, \times 10^{-12} \, \mu_{\rm B}\,,
\label{eq:mu_lim_summ}
\end{eqnarray}
depending on the explosion model. 

%%%%%%%%%%%%%%%%%%%%%%%%%%%%%%%%%%%%%%%%%%%%%%%%%%%%%%%%%%%%%%%%%%%%%%
\section{Conclusions}
\label{sec:Conclusions}
%%%%%%%%%%%%%%%%%%%%%%%%%%%%%%%%%%%%%%%%%%%%%%%%%%%%%%%%%%%%%%%%%%%%%%

The right-handed neutrino luminosity caused by the neutrino helicity-flip process 
under the conditions of the supernova core, 
where the produced left-handed neutrinos could convert due to the neutrino magnetic 
moment interaction into the right-handed neutrinos, being sterile with respect to the weak 
interaction, is reinvestigated. 
Instead of the uniform ball model for the SN core used in previous 
analyses, realistic models for radial distributions and time evolution of physical parameters in 
the SN core are considered. The upper bounds on the flavor- and time-averaged 
magnetic moment of the Dirac type neutrino are obtained in those models, 
from the condition of not-affecting the total cooling time scale 
significantly:  
\begin{eqnarray}
\bar \mu_\nu < (1.1 - 2.7) \, \times 10^{-12} \, \mu_{\rm B}\,,
\label{eq:mu_lim_summ2}
\end{eqnarray}
depending on the explosion model. 

In the recent paper,\cite{Lychkovskiy:2009} the sterile right-handed neutrino 
luminosity was calculated with taking account of radial distributions of the 
supernova core parameters, using the one-dimensional astrophysical 
code ``Boom''.\cite{Boom} The supernova matter state parameters were 
calculated as the functions of coordinate and time during 250 ms after 
the bounce. We should give a comment on Table 2 of 
Ref.~\refcite{Lychkovskiy:2009}. At first glance, the result for the 
right-handed neutrino luminosity, (0.5--1.1) $\times 10^{50}$ erg/s, 
obtained in Ref.~\refcite{Lychkovskiy:2009} is in agreement with the result 
(0.4--4) $\times 10^{50}$ erg/s by R. Barbieri and 
R.\,N. Mohapatra,\cite{Barbieri:1988} and contradicts to 
our\cite{Kuznetsov:2007} result (3.8--22) $\times 10^{50}$ erg/s. 
However, as is seen from the paper, the result of Ref.~\refcite{Lychkovskiy:2009} 
for the minimal value of the right-handed neutrino luminosity, 0.5 $\times 10^{50}$ erg/s 
corresponds to the effective temperature of the emitting matter $T_{eff} \simeq$ 10 MeV, 
while the minimal values of the luminosities of Refs.~\refcite{Barbieri:1988} 
and~\refcite{Kuznetsov:2007} correspond to the averaged temperature 30 MeV. 
Taking into account the strong dependence of the right-handed neutrino luminosity 
on the temperature, mentioned in Refs.~\refcite{Kuznetsov:2007,Kuznetsov:2009}, 
one should conclude that the results of Refs.~\refcite{Lychkovskiy:2009} 
and~\refcite{Kuznetsov:2007} are in agreement, and both contradict to the 
result of Ref.~\refcite{Barbieri:1988}. 

%\newpage

%%%%%%%%%%%%%%%%%%%%%%%%%%%%%%%%%%%%%%%%%%%%%%%%%%%%%%%%%%%%%%%%%%%%%%
\section*{Acknowledgements}
%%%%%%%%%%%%%%%%%%%%%%%%%%%%%%%%%%%%%%%%%%%%%%%%%%%%%%%%%%%%%%%%%%%%%%

We are grateful to Hans-Thomas Janka and Bernhard M\"uller for 
providing us with detailed data on radial distributions and time evolution 
of physical parameters in the supernova core, obtained in their 
model of the SN explosion. We thank Oleg Lychkovskiy for useful 
discussion. 
 
This work was supported by the 
the Council of the President of the Russian Federation 
for Support of Young Scientists and Leading Scientific Schools 
(project No. NSh-497.2008.2), the Ministry of Education 
and Science of the Russian Federation (Program 
``Development of the Scientific Potential of the Higher 
Education'', project No. 2.1.1/510), and the Russian Foundation 
for Basic Research (project No. 07-02-00285a). 

%%%%%%%%%%%%%%%%%%%%%%%%%%%%%%%%%%%%%%%%%%%%%%%%%%%%%%%%%%%%%%%%%%%%%%

\end{document}